\documentclass[pdflatex]{article}%

\usepackage{graphicx}%
\usepackage{multirow}%
\usepackage{amsmath,amssymb,amsfonts}%
\usepackage{amsthm}%
\usepackage{mathrsfs}%
\usepackage[title]{appendix}%
\usepackage{xcolor}%
\usepackage{textcomp}%
\usepackage{manyfoot}%
\usepackage{booktabs}%
\usepackage{algorithm}%
\usepackage{algorithmicx}%
\usepackage{algpseudocode}%
\usepackage{listings}%
\usepackage{soul}%
\usepackage{authblk}%
\usepackage[a4paper, total={6in, 8in}]{geometry}

\theoremstyle{thmstyleone}%
\theoremstyle{thmstyletwo}%

\theoremstyle{thmstylethree}%

\begin{document}

\title{Wetting effects on the dynamics of droplets and bubbles at surfaces}

\author[1,*]{Yifan Han}

\author[1,2,3]{Kerstin Eckert}

\author[1]{Gerd Mutschke}

\affil[1]{Institute of Fluid Dynamics, Helmholtz-Zentrum Dresden-Rossendorf, Bautzner Landstrasse 400, 01328 Dresden, Germany}

\affil[2]{Institute of Process Engineering and Environmental Technology, Technische Universit\"at Dresden, 01062 Dresden, Germany}

\affil[3]{Hydrogen Lab, School of Engineering, Technische Universit\"at Dresden, 01062 Dresden, Germany}

\affil[*]{Email: yi.han@hzdr.de}

\date{June 30, 2026}

\maketitle


\abstract{
Dynamic wetting plays a fundamental role in the dynamics of droplets and bubbles at solid surfaces by influencing contact line motion and interfacial evolution. 
In this work, three representative wetting-controlled benchmarks, namely droplet splashing, bubble coalescence at solid surfaces, and bubble dynamics under shear flow, are investigated using a three-dimensional volume-of-fluid framework coupled with a recently developed dynamic wetting model based on contact line velocity reconstruction method~\cite{han2026geometric}.
The model is first validated against experimental observations and literature data for droplet splashing and bubble coalescence. It accurately reproduces the transient contact line evolution, splashing morphology, and coalescence dynamics. In particular, dynamic wetting suppresses the premature bubble detachment predicted by static wetting models and yields substantially improved agreement with experimental observations. 
In shear flow, contact angle hysteresis and contact line dissipation strongly influence bubble deformation, sliding, and detachment. 
These results demonstrate that accurate treatment of dynamic wetting is essential for predicting wetting-controlled droplets and bubbles involving rapid contact line motion, strong interfacial deformation, and topology changes.
}




\maketitle

\section{Introduction}
\label{intro}

Wetting phenomena are ubiquitous in multiphase systems involving droplets and bubbles interacting with solid surfaces and play a fundamental role in interfacial transport processes. 
They arise in a wide range of natural processes and engineering applications, including raindrop sliding, inkjet printing, boiling, electrochemical gas evolution, coating technologies, and microfluidic applications\cite{de1985wetting,bonn2009wetting,snoeijer2013moving}.
In these systems, the motion of the three-phase contact line governs the local balance between capillary, viscous, and inertial forces, 
thereby controlling interfacial deformation, spreading, recoiling, sliding, coalescence, and detachment.
As a result, the wetting dynamics strongly influences the morphology, mobility, and lifetime of droplets and bubbles at solid surfaces, ultimately affecting momentum, heat and mass transfer across a broad range of spatial and temporal scales. 
An accurate understanding and prediction of the wetting behavior is therefore essential for optimizing the performance of many multiphase transport processes and energy-conversion technologies.

Unlike equilibrium wetting, most practical multiphase flows involve rapidly evolving interfaces subjected to competing capillary, viscous, inertial, gravitational, and externally imposed forcing effects \cite{blake2006physics}. Under such conditions, the apparent contact angle becomes intrinsically coupled to the local contact line motion and may deviate substantially from its equilibrium value $\theta_s$. 
The resulting advancing and receding dynamics, together with contact angle hysteresis (CAH), contact line friction, and pinning/depinning events, govern the dissipation mechanisms at the three-phase contact line and determine the dynamic evolution of the apparent contact angle during wetting and dewetting processes \cite{butt2022contact,barrio2020contact,demirkir2024life,demirkir2025jump}.
Consequently, dynamic wetting not only controls the evolution of droplets and bubbles on solid surfaces but also determines the onset and development of interfacial instabilities such as splashing\cite{josserand2016drop}, breakup, and shear-induced detachment \cite{kossolapov2024bubble}. 
A precise representation of dynamic wetting is therefore essential for predictive modeling of multiphase flows involving moving contact lines and strong interface deformation.

The accurate prediction of these phenomena requires numerical models capable of resolving both interface evolution and moving contact lines. However, implementing dynamic wetting in multiphase-flow simulations remains challenging because the apparent contact angle depends sensitively on the local contact line motion, which is often difficult to implement robustly especially
in rapidly evolving flows and in cases
of contact angle hysteresis.
Various approaches to model
dynamic wetting
have been implemented in
phase-field, level-set and volume-of-fluid (VOF) methods~\cite{ding2007wetting,yokoi2009numerical,ZHANG2020109636,chen2025extended,chen2025volume}.
Among the available approaches, the VOF method is particularly attractive owing to its excellent mass conservation properties and its ability to capture large interface deformation and morphology changes such as
breakup and coalescence \cite{popinet2018numerical}. 
Substantial efforts have been devoted to the incorporation of dynamic wetting into VOF frameworks. 
However, most implementations are available only for two-dimensional geometries which limits their application \cite{han2025numerical, huang20252d}.
Recent developments 
for three-dimensional problems
include a Hoffman-function-based dynamic contact angle model to simulate droplet impact \cite{vozhakov2025numerical} and conservative sharp VOF formulation capable of describing dynamic wetting 
with contact angle hysteresis
on complex boundaries \cite{huang20263d}. 
Despite these advances, accurately coupling dynamic wetting with interface evolution remains a challenge in strongly deforming three-dimensional cases. 
In particular, the dynamic contact angle depends critically on the local contact line velocity, whose evaluation can become sensitive to interface deformation, mesh resolution, and transient topological changes~\cite{DUPONT20102453,han2025numerical}. 
As a result, uncertainties in contact line reconstruction can significantly affect the prediction of wetting dynamics and interfacial evolution.

To address the challenges associated with moving contact lines in complex multiphase flows, we recently developed a geometric interpolation framework for dynamic wetting in three-dimensional VOF simulations \cite{han2026geometric}. 
Building upon this methodology, the present work investigates the effects of dynamic wetting on three representative wetting-controlled interfacial flows: droplet splashing, coalescence-induced bubble dynamics at solid surfaces, and bubble deformation and detachment under shear flow. 
By examining these systems, which involve rapid contact line motion, strong interface deformation, and topology changes, we aim to elucidate how dynamic wetting influences interfacial evolution, instability development, and detachment behavior in complex droplet and bubble systems.

The remainder of this paper is organized as follows. Section~\ref{method} introduces the numerical methodology and the dynamic wetting framework. Section~\ref{results} presents the results for droplet splashing, bubble coalescence, and bubble dynamics under shear flow. Finally, the main conclusions are given in Section~\ref{conclu}.

\section{Numerical method}
\label{method}

The dynamics of droplets and bubbles that interact with solid surfaces is simulated using the open-source finite-volume solver \textsc{Basilisk} \cite{popinet2009accurate}. 
The numerical framework combines a geometric volume-of-fluid (VOF) method with adaptive mesh refinement (AMR) and a dynamic wetting model to resolve moving contact lines in three-dimensional multiphase flows.

\subsection{Governing equations}
\label{method:eqs}
The incompressible two-phase flow is governed by the Navier--Stokes equations.
\begin{equation}
\nabla \cdot \mathbf{u} = 0,
\label{eq:continuity}
\end{equation}

\begin{equation}
\rho
\left(
\frac{\partial \mathbf{u}}{\partial t}
+
\mathbf{u}\cdot\nabla\mathbf{u}
\right)
=
-\nabla p
+
\nabla\cdot
\left[
\mu
\left(
\nabla \mathbf{u}
+
\nabla \mathbf{u}^{T}
\right)
\right]
+
\gamma \kappa \delta_s \mathbf{n}
+
\rho \mathbf{g},
\label{eq:NS}
\end{equation}
\noindent
where $\mathbf{u}=(u_x,u_y,u_z)$ denotes the phase-averaged fluid velocity in the Cartesian coordinate system,
$p$ the pressure, $\rho$ and $\mu$ the phase-dependent density and viscosity, $\gamma$ the surface tension coefficient, $\kappa$ the interface curvature, and $\mathbf{g}$ the gravitational acceleration. The surface tension force is implemented using the balanced-force continuum surface force (CSF) formulation, which significantly reduces parasitic currents near the interface.

The phase distribution is represented by the VOF scalar $f$, where $f=1$ and $f=0$ correspond to the liquid and gas phases, respectively. The transport of the interface is governed by
\begin{equation}
\frac{\partial f}{\partial t}
+
\nabla \cdot (f \mathbf{u}) = 0.
\label{eq:VOF}
\end{equation}

Fluid properties are interpolated according to the local volume fraction,
\begin{equation}
\rho = f \rho_l + (1-f)\rho_g,
\qquad
\mu = f \mu_l + (1-f)\mu_g,
\end{equation}
\noindent
where the subscripts $l$ and $g$ denote the liquid and gas phases, respectively.

The liquid--gas interface is captured using a geometric volume-of-fluid (VOF) method based on the piecewise-linear interface calculation (PLIC) reconstruction implemented in \textsc{Basilisk} \cite{popinet2009accurate}. The interface curvature is evaluated using the height-function formulation, which provides second-order accuracy and robust surface-tension calculations for capillary-dominated flows. 
Adaptive mesh refinement is applied using wavelet-based error estimates of the phase fraction and velocity fields, with refinement concentrated near the interface and the three-phase contact line. 
The resulting octree grid efficiently captures multiscale interfacial structures in three dimensions. 
Time integration is performed explicitly with a Courant number less than 0.5. The maximum time step is also limited to $\Delta t_{\max}=10^{-7}\,\mathrm{s}$ to ensure accurate resolution of the capillary-wave propagation and the motion of the contact line.

\subsection{Dynamic wetting model}

The dynamics of the contact line is modeled using the geometric dynamic wetting framework developed in our previous work \cite{han2026geometric}. 
Existing VOF-based dynamic wetting approaches commonly evaluate the apparent contact angle from the
contact line velocity estimated using local flow information \cite{vozhakov2025numerical}. 
Although these methods have demonstrated good predictive capability for a variety of wetting problems, the estimation of the contact line velocity may become sensitive to local interface deformation in rapidly evolving three-dimensional flows. 
The present framework incorporates a geometric interpolation procedure that reconstructs the contact line velocity directly from the evolution of local interface near the wall. 
This treatment improves the consistency between interface motion and contact line kinematics, providing a robust coupling between the hydrodynamic solution and dynamic wetting model. 
The apparent contact angle is defined and measured on the liquid side.
The reconstructed contact line velocity is then used to evaluate the apparent dynamic contact angle.

The apparent dynamic contact angle, $\theta_{\mathrm{app}}$, is calculated using the dynamic wetting model proposed by Dwivedi et al.~\cite{dwivedi2022dynamic},
\begin{equation}
\theta_{\mathrm{app}}^{3}
=
\left\{
\arccos
\left[
\cos \theta_s
-
\frac{\xi \mathrm{Ca}}{\mu}
-
\frac{C_{\mathrm{pin}}
\tanh(C\mathrm{Ca})}{\gamma}
\right]
\right\}^{3}
+
9\mathrm{Ca}\ln(\epsilon),
\label{eq:DCA}
\end{equation}
\noindent
where $\theta_s$ is the equilibrium contact angle, $\mathrm{Ca}=\mu u_{cl}/\gamma$ is the capillary number based on the contact line velocity $u_{cl}$, $\xi$ is the contact line friction coefficient, and $\epsilon$ is the microscopic cutoff length appearing in the hydrodynamic description of moving contact lines.
The coefficient $C$ controls the activation of the pinning force and determines how rapidly contact-angle hysteresis develops with increasing capillary number. Larger values of $C$ lead to an earlier saturation of the hysteresis contribution.
The pinning coefficient $C_{\mathrm{pin}}$ accounts for contact-angle hysteresis and is defined as
\begin{equation}
C_{\mathrm{pin}}
=
\begin{cases}
\gamma \left(\cos\theta_s-\cos\theta_a\right),
& \mathrm{Ca}>0,\\[6pt]
\gamma \left(\cos\theta_r-\cos\theta_s\right),
& \mathrm{Ca}<0,
\end{cases}
\label{eq:Cpin}
\end{equation}
\noindent
where $\theta_a$ and $\theta_r$ denote the advancing and receding contact angles, respectively.

The reconstructed dynamic contact angle is imposed through the height-function framework, which provides a geometrically consistent implementation of contact angle boundary conditions in VOF simulations \cite{afkhami2004height,afkhami2008height,han2021consistent}. 
This treatment ensures a consistent coupling between interface reconstruction, capillary forces, and dynamic wetting, enabling accurate simulation of advancing and receding contact line motion, contact-angle hysteresis, and transient pinning/depinning phenomena in complex three-dimensional flows. 
A distinctive feature of the present framework is the geometric interpolation of the contact line velocity directly from the local interface evolution, providing a direct link between interface kinematics and dynamic wetting. 
The numerical procedure is described in the following section.

\subsection{Geometric reconstruction of contact line velocity}

Unlike conventional VOF implementations, where the contact line velocity is estimated from local or averaged velocity fields near the wall, the present approach reconstructs it directly from the evolution of the geometrically reconstructed interface. 
This treatment reduces the sensitivity of the evaluated contact line velocity to mesh resolution, local velocity fluctuations, and severe interface deformation in rapidly evolving three-dimensional flows.

\begin{figure}
    \centering
    \includegraphics[width=0.5\linewidth]{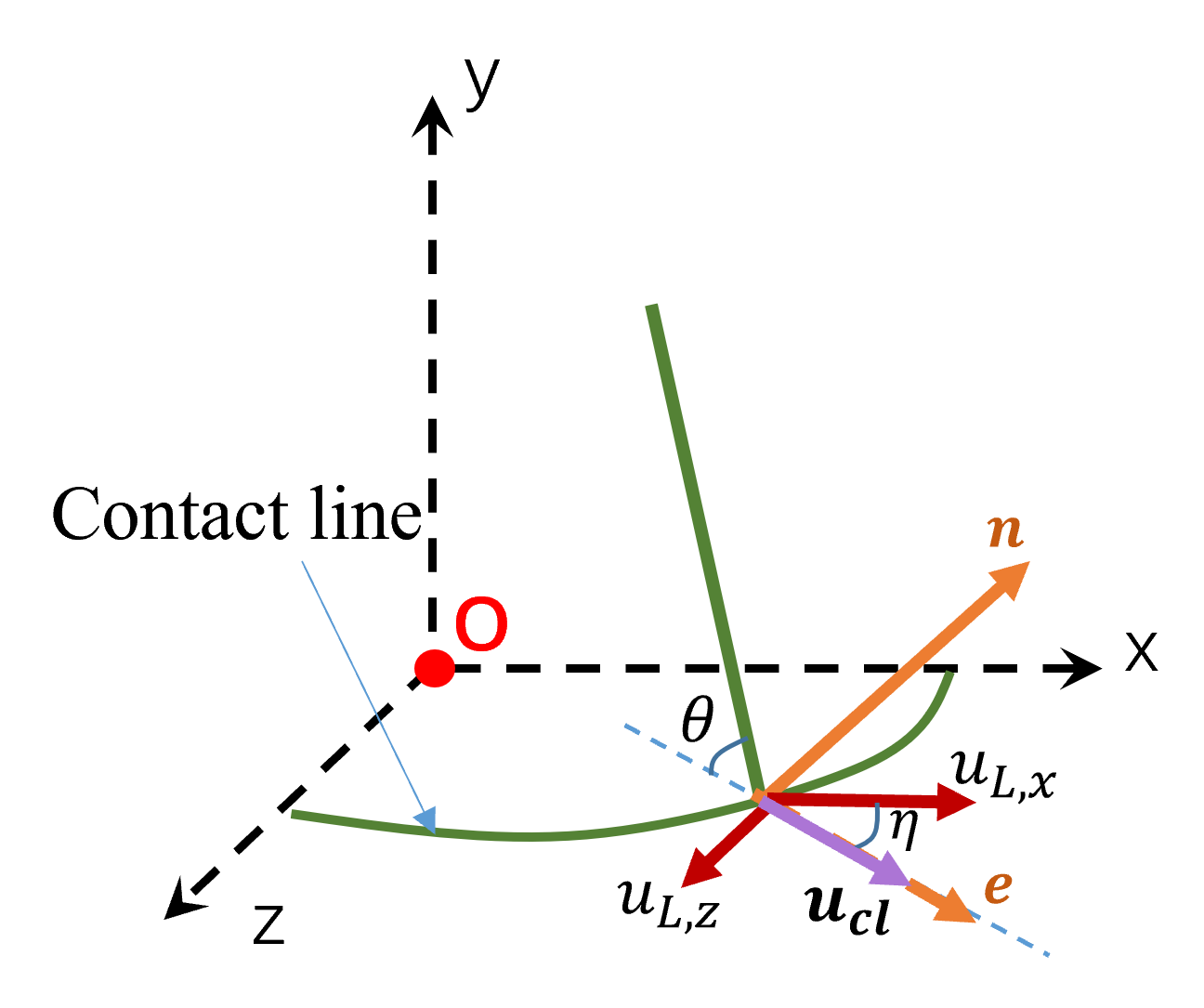}
    \caption{Geometric reconstruction of the contact-line velocity near a moving contact line.}
    \label{fig:Fig_0_sketch_ucl}
\end{figure}

The displacement of the local interface is first obtained from the PLIC-reconstructed interface in neighboring wall-adjacent cells. 
A geometric interpolation procedure is then employed to determine the interfacial velocity vector $\mathbf{u}_{\Sigma}$ in the vicinity of the contact line, as illustrated in Fig.~\ref{fig:Fig_0_sketch_ucl}. 
Since only the normal component of the interfacial motion contributes to contact line displacement, the contact line velocity is computed by projecting the reconstructed interfacial velocity onto the local contact line direction
\begin{equation}
\mathbf{u}_{cl} = (u_{L,x}\cos\eta + u_{L,z}\sin\eta)\mathbf{e},
\label{eq06}
\end{equation}
\noindent
where $\mathbf{e}$ denotes the unit tangent vector along the contact line, $\eta$ is the local contact line orientation angle, and $u_{L,x}$ and $u_{L,z}$ are the geometrically reconstructed interfacial velocity components in the streamwise and spanwise directions, respectively. 
The projection extracts the component of the reconstructed interfacial velocity responsible for contact line motion while filtering the normal component associated with interface deformation.
By reconstructing the contact line velocity directly from interface evolution, the method establishes a direct and physically consistent coupling between contact line kinematics and interface motion than approaches based solely on local flow quantities. As demonstrated in our previous work \cite{han2026geometric}, this treatment substantially improves the robustness and accuracy of contact line tracking in three-dimensional flows involving strong interface deformation, splashing, coalescence, and detachment.

The reconstructed contact line velocity is subsequently used to evaluate the capillary number in Eq.~(\ref{eq:DCA}), thereby establishing a direct coupling between interface evolution and the dynamic wetting model. This coupling enables consistent prediction of the apparent angle evolution throughout complex wetting and dewetting processes. Further details of the geometric interpolation procedure, numerical implementation, validation tests, and convergence studies can be found in Ref.~\cite{han2026geometric}.

\section{Results and discussion}
\label{results}

The predictive capability of the present dynamic wetting framework is assessed using three representative wetting-controlled multiphase flows: droplet splashing during impact, coalescence-induced bubble dynamics at solid surfaces, and bubble deformation and detachment under shear flow.
These benchmark problems involve rapid contact line motion, strong capillary-driven deformation, and topology changes, thereby providing stringent tests for assessing the performance of the dynamic wetting framework.
Unless otherwise specified, all simulations consider an air--water system under ambient conditions with $\rho_l=998~\mathrm{kg,m^{-3}}$, $\rho_g=1.2~\mathrm{kg,m^{-3}}$, $\mu_l=1.0\times10^{-3}~\mathrm{Pa,s}$, $\mu_g=1.8\times10^{-5}~\mathrm{Pa,s}$, and $\gamma=0.072~\mathrm{N,m^{-1}}$. 
The dynamic wetting framework described in Section~\ref{method} is employed throughout, whereas the corresponding wetting parameters are specified separately for each benchmark case. 
All simulations are performed in parallel using up to 128 CPU cores, with computational costs ranging from several hundred to several thousand CPU-hours.

\subsection{Droplet splashing at surfaces}
\label{results:splashing}

Droplet splashing represents one of the most demanding benchmarks for dynamic wetting models because it involves rapid contact line motion, strong capillary--inertial interactions, and severe interface deformation accompanied by topology changes. 
Extensive experimental and numerical studies have shown that contact line dynamics strongly affect spreading, lamella formation, rim instability, and splashing onset \cite{josserand2016drop,pasandideh1996capillary,vozhakov2025numerical}.
Accurate prediction of these processes therefore requires a robust coupling between interface evolution and dynamic wetting.
To assess the performance of the present framework under highly transient conditions, droplet impact at $\mathrm{We}=260$ is studied numerically based on the experimental configuration of Vozhakov et al.~\cite{vozhakov2025numerical}. 
The predictions are compared with experimental measurements and the DCAH model reported in the literature, providing a stringent validation of the geometric dynamic wetting framework under splashing conditions.

\begin{figure}
\centering
\begin{minipage}[b]{0.47\textwidth}
    \centering
    \includegraphics[width=\textwidth]{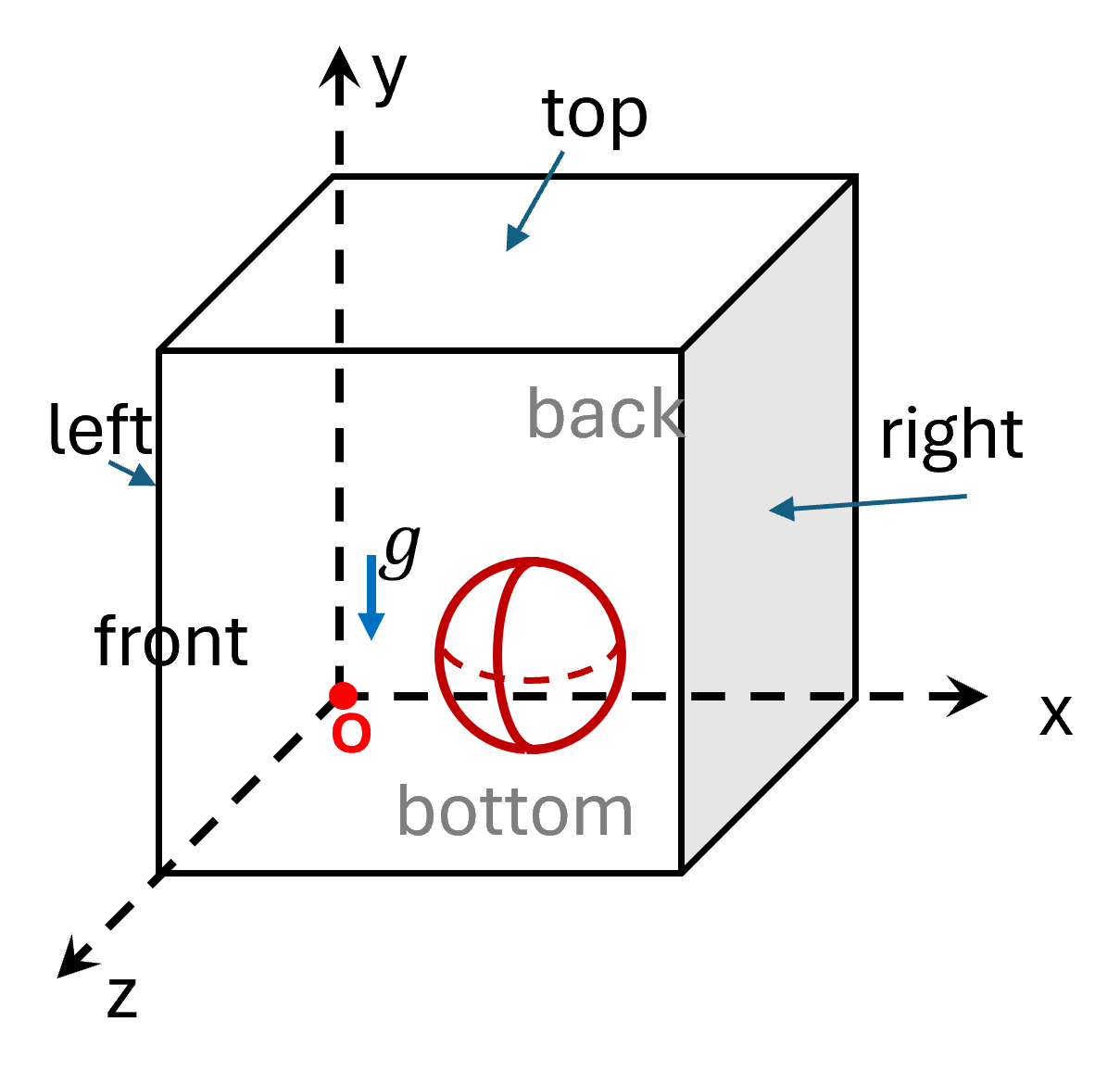}\\
    \textbf{(a)}
\end{minipage}
\hfill
\begin{minipage}[b]{0.47\textwidth}
    \centering
    \includegraphics[width=\textwidth]{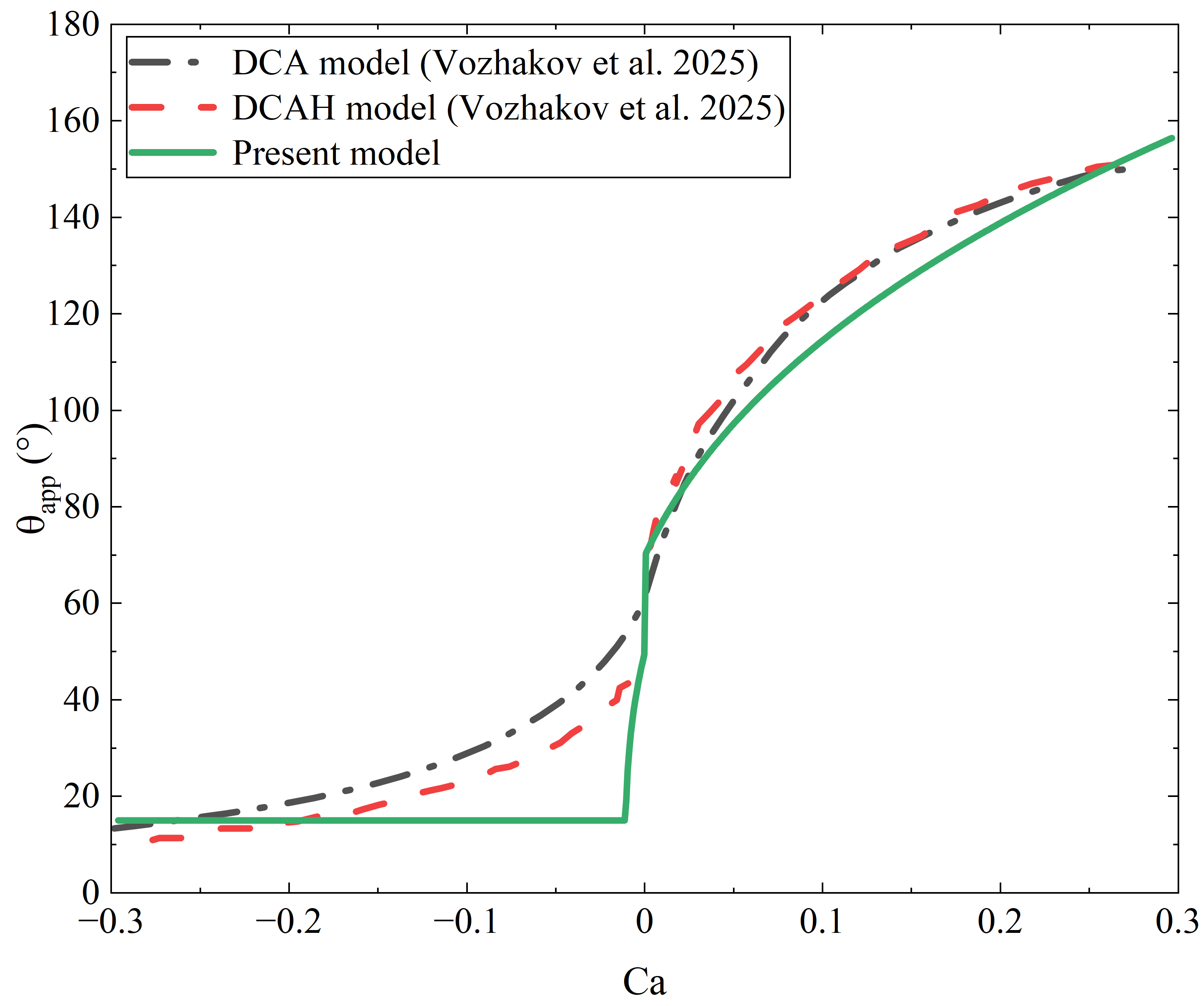}\\
    \textbf{(b)}
\end{minipage}
\caption{
Numerical setup and dynamic wetting characterization for droplet splashing.
(a) Computational domain and initial droplet configuration.
(b) Apparent contact angle predicted by different wetting models as a function of capillary number.
}
\label{fig-splashing-setup}
\end{figure}

The computational configuration is illustrated in Fig.~\ref{fig-splashing-setup}(a). To enable a direct assessment of the dynamic wetting model, the droplet-impact problem is reproduced using the same physical properties, geometric parameters, computational domain, and impact conditions as those reported by Vozhakov et al.~\cite{vozhakov2025numerical}. 
In particular, the initial droplet diameter is $D_0=2.1~\mathrm{mm}$, the impact velocity is $U_0=3.0~\mathrm{m,s^{-1}}$, corresponding to $\mathrm{We}=260$, and the fluid properties correspond to the air--water system at ambient conditions. 
The numerical setup, boundary conditions, and mesh refinement strategy are also kept identical to those used in the reference study. 
The only difference lies in the treatment of dynamic wetting. 
Whereas Vozhakov et al.~\cite{vozhakov2025numerical} employed a Hoffman-function-based DCA and DCAH model, the present simulations adopt the geometric dynamic wetting framework described in Section~\ref{method}. The dynamic wetting parameters used in the present study are summarized in Table~\ref{tab-splash}. 
During the late-stage recoil, the dynamic wetting model may predict strong dewetting accompanied by rapid recession of the contact line. 
To prevent unrealistically small apparent contact angles and to account for the finite wettability that persists near the contact line, a minimum apparent contact angle of $\theta_{\min}=15^\circ$ is prescribed. 
The constraint is only activated under strong dewetting conditions and has a negligible influence on the spreading stage of the impact process.

    \begin{table}[]
    \centering
    \caption{\label{tab-splash}Dynamic wetting parameters of a TMCS-treated glass surface with $\theta_s=35^\circ$.}
    \begin{tabular}{llllll}
    \hline
             $\theta_s (\mathrm{^\circ})$    &$\theta_a  (\mathrm{^\circ})$  &$\theta_r  (\mathrm{^\circ})$ & $\xi  (\mathrm{Pa \cdot s})$ & \quad $C$ & \quad$\epsilon$ \\\hline
            60 & 70 & 50 & \quad 0.001 & $2\times 10^4$ & $1.0\times 10^4$    \\\hline
    \end{tabular}
    \end{table}

\begin{figure}
\centering
\begin{minipage}{0.95\textwidth}
    \centering
    \includegraphics[width=\textwidth]{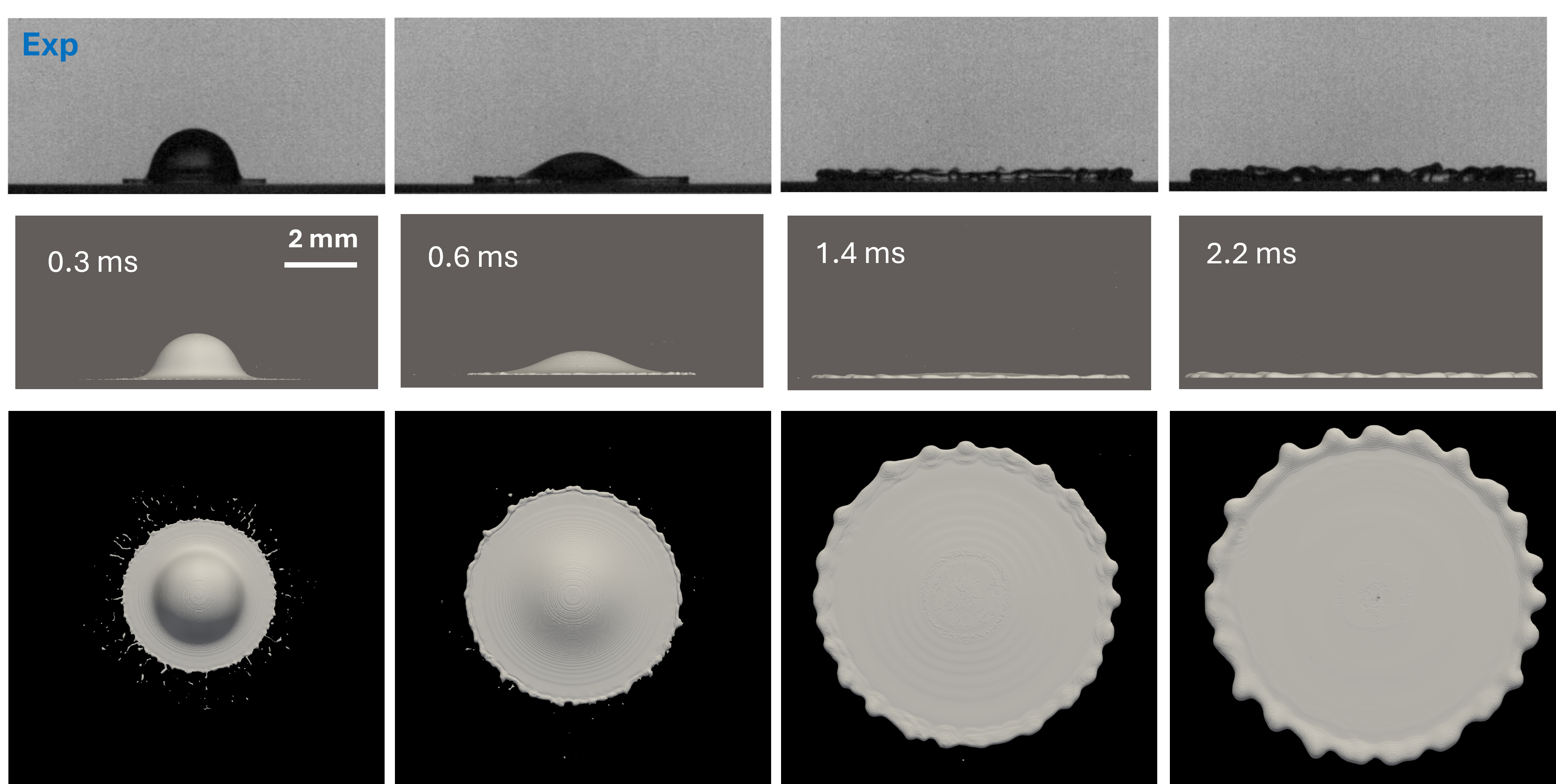}
    \label{fig-a}
    \textbf{(a)}
\end{minipage}

\vspace{0.3cm}
\begin{minipage}{0.48\textwidth}
    \centering
    \includegraphics[width=\textwidth]{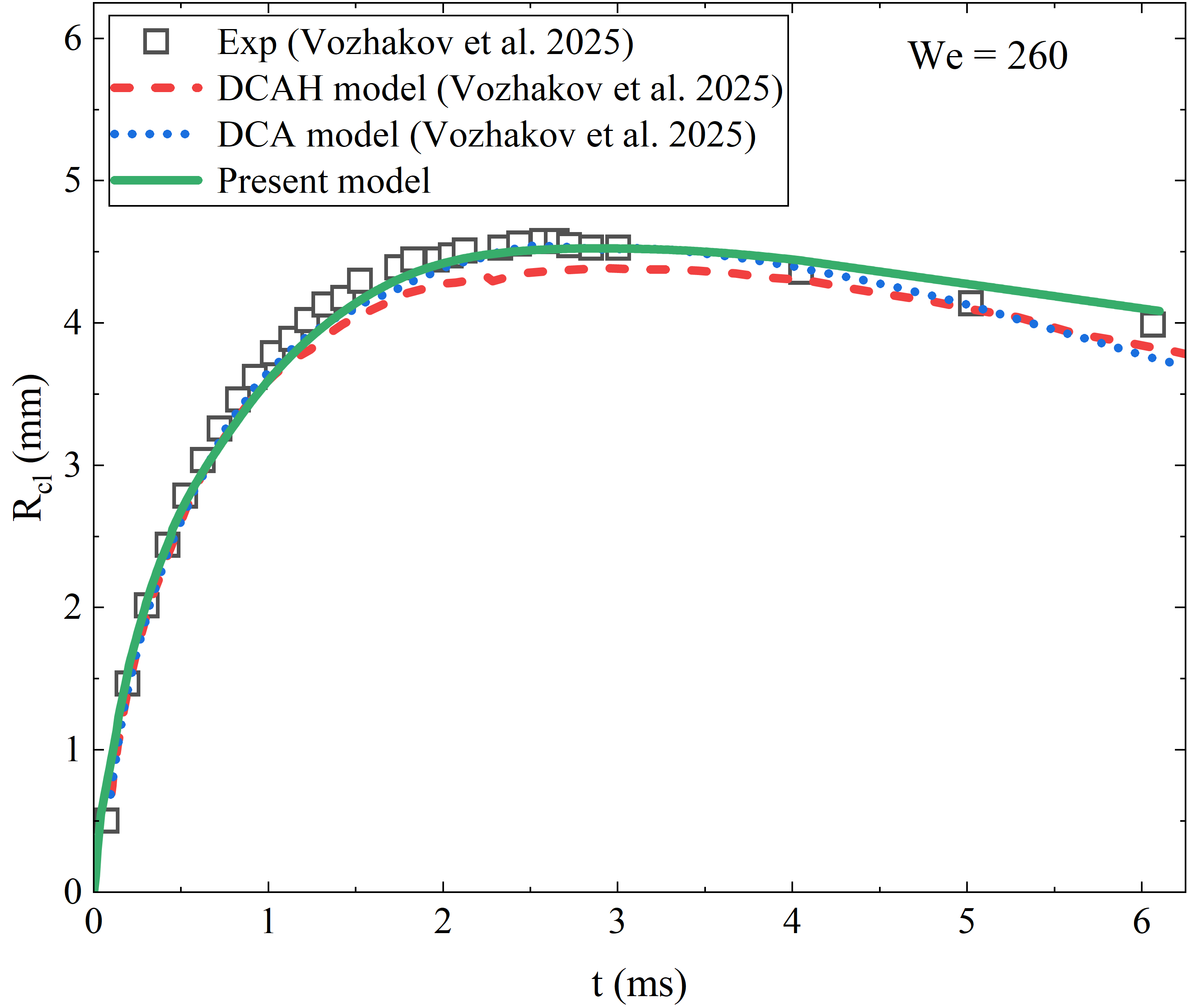}
    \label{fig-b}
    \textbf{(b)}
\end{minipage}
\hfill
\begin{minipage}{0.48\textwidth}
    \centering
    \includegraphics[width=\textwidth]{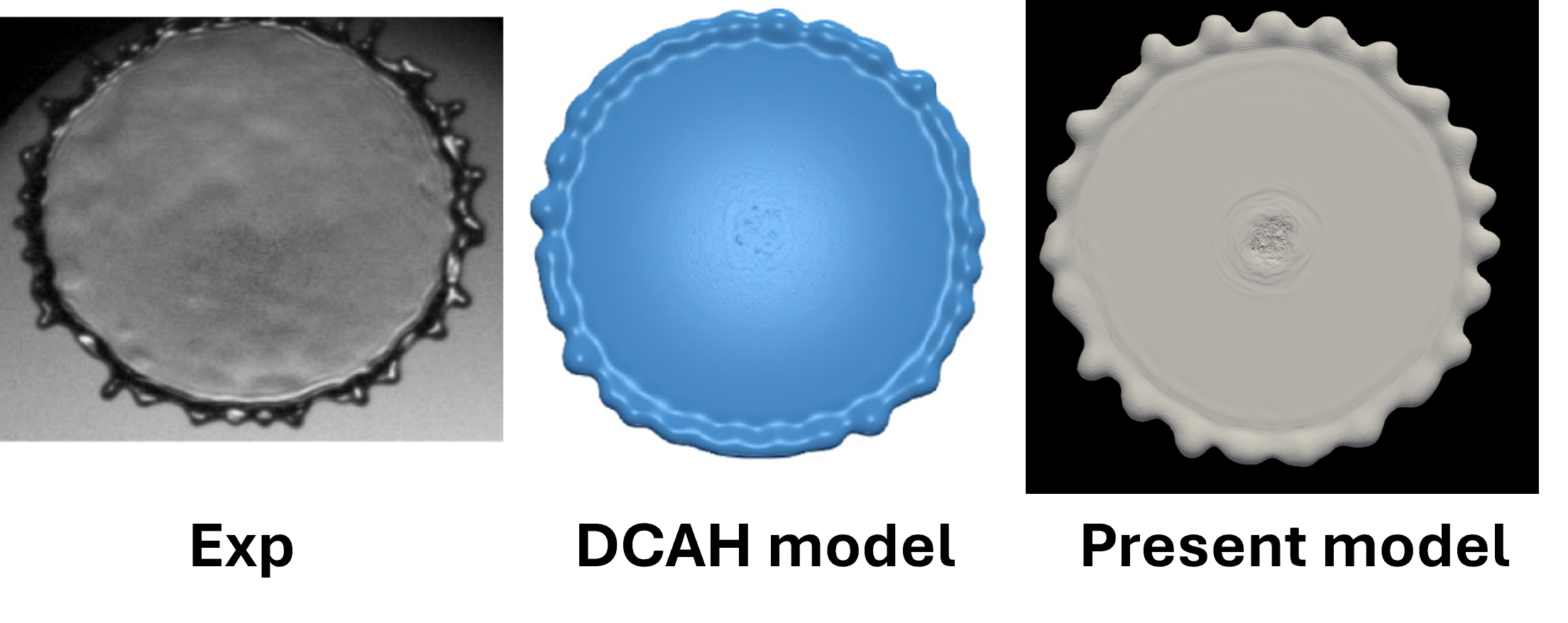}
    \label{fig-c}
    \textbf{(c)}
\end{minipage}
\caption{
Comparison of droplet splashing dynamics at $\mathrm{We}=260$ between experiment and numerical simulation.
(a) Temporal evolution of droplet impact and splashing. (Top: experiment)
(b) Time evolution of the contact line radius $R_{cl}$.
(c) Comparison of the maximum spreading morphology.
}
\label{fig:splashing_validation}
\end{figure}

Figure~\ref{fig:splashing_validation} presents a comprehensive comparison of the droplet splashing dynamics between the experiment of Vozhakov et al.~\cite{vozhakov2025numerical} and the present simulations at $\mathrm{We}=260$. 
Figure~\ref{fig:splashing_validation}(a) compares the temporal evolution of the impact process. 
The numerical results successfully reproduce the major stages of splashing, including the initial spreading, thin-lamella formation, rim growth, and subsequent crown development. Good agreement with the experimental observations is obtained throughout the impact process. 
In particular, the top-view morphology reveals the emergence of azimuthal rim corrugations along the expanding contact line, which subsequently evolve into the characteristic crown-like splashing structure.

To quantitatively assess the wetting dynamics, Fig.~\ref{fig:splashing_validation}(b) compares the evolution of the contact line radius $R_{cl}$ with the experimental measurements and the DCA/DCAH model predictions reported by Vozhakov et al.~\cite{vozhakov2025numerical}. 
The present model closely follows the experimental data during both the spreading and recoiling stages and shows improved agreement compared with the DCA and DCAH models, particularly near the maximum spreading state where wetting dissipation and capillary forces strongly influence the contact line motion. 
This improvement indicates that resolving the contact line motion consistently with an accurate dynamic wetting model is important for capturing the rapid spreading--recoiling transition under splashing conditions.
Further insight is provided by the comparison of the maximum spreading morphology shown in Fig.~\ref{fig:splashing_validation}(c). 
While the DCAH model predicts a relatively smooth and nearly axisymmetric rim, it underestimates the development of azimuthal corrugations observed experimentally. 
In contrast, the present model reproduces both the amplitude and wavelength of the rim undulations more accurately, resulting in substantially better agreement with the experimental splashing pattern. 

\subsection{Bubble coalescence at surfaces}
\label{results:coalescence}

Coalescence-induced bubble dynamics provides another representative example of wetting-controlled interfacial flows involving rapid contact line motion, strong capillary--inertial recoil, and topology changes. 
Such processes are encountered in boiling, electrochemical gas evolution, and multiphase transport systems, where bubble coalescence frequently determines gas removal efficiency and surface coverage. 
In this subsection, the capability of the present dynamic wetting framework to predict coalescence-induced bubble dynamics is assessed through direct comparison with the experiments of Wang et al.~\cite{wang2025role}.

\begin{figure}
\centering
\begin{minipage}[b]{0.49\textwidth}
    \centering
    \includegraphics[width=\textwidth]{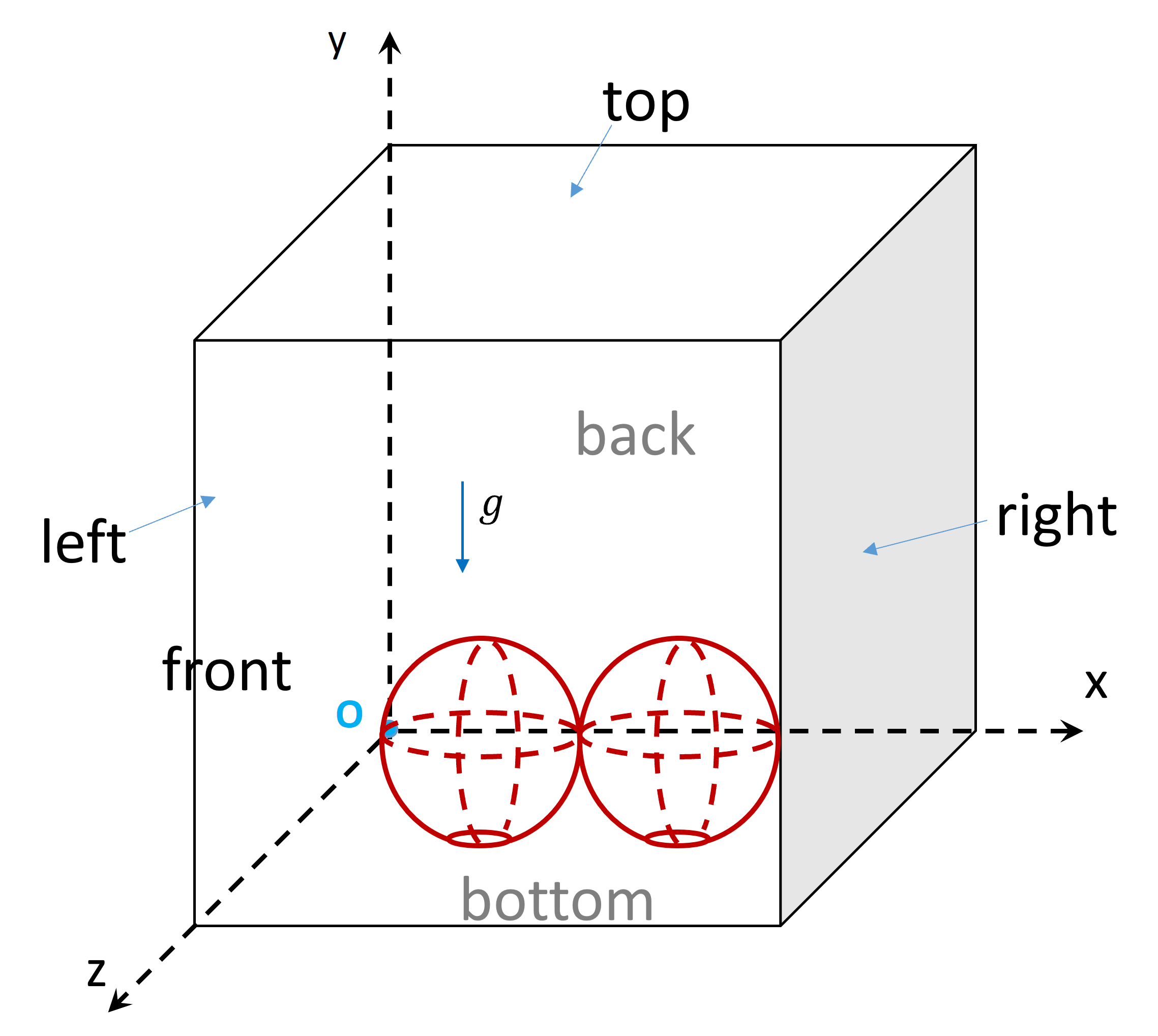}\\
    \textbf{(a)}
\end{minipage}
\hfill
\begin{minipage}[b]{0.45\textwidth}
    \centering
    \includegraphics[width=\textwidth]{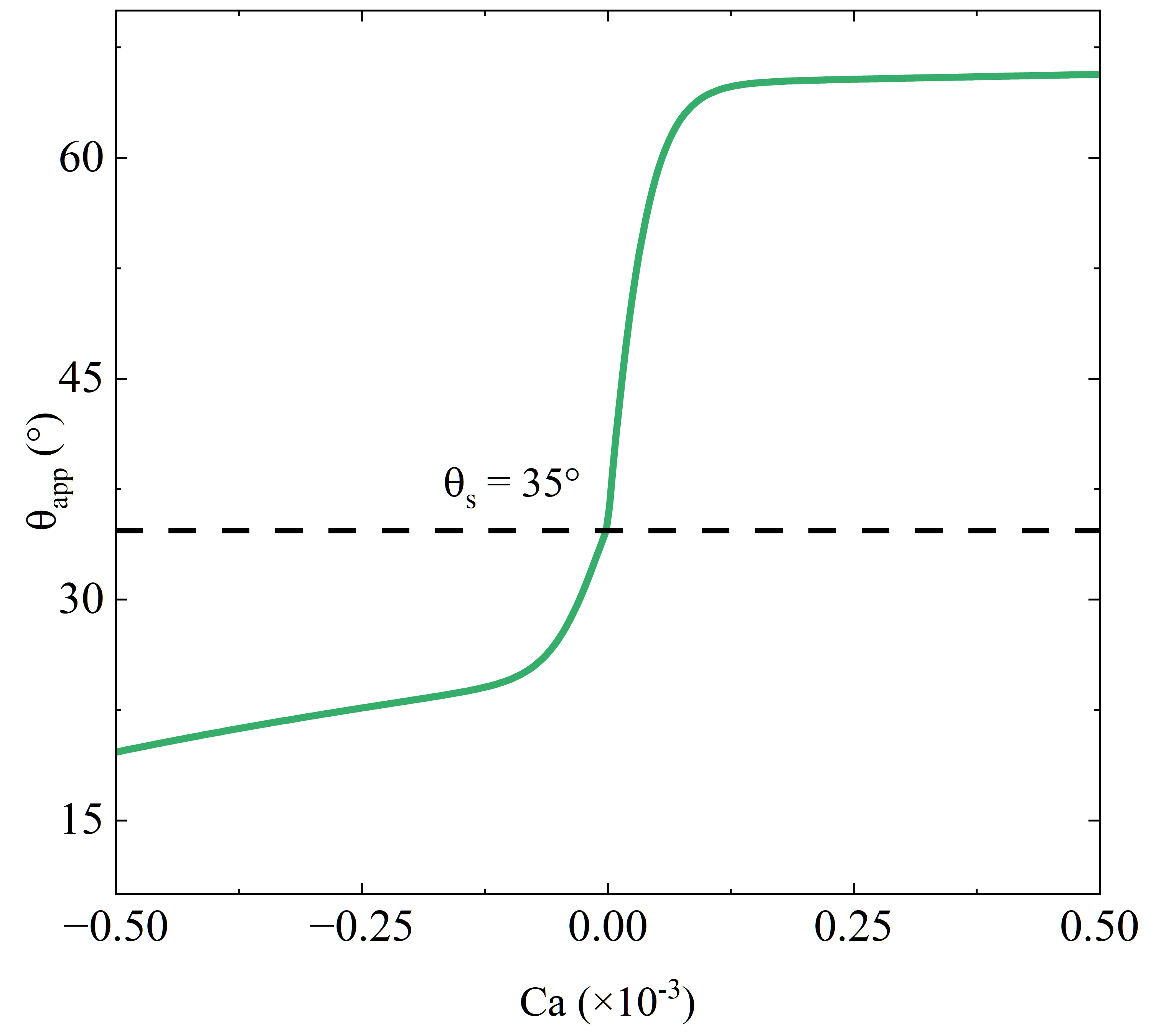}\\
    \textbf{(b)}
\end{minipage}
\caption{
Numerical setup and dynamic wetting characterization for bubble coalescence.
(a) Computational domain and initial configuration of two identical bubbles attached to a solid surface.
(b) Apparent contact angle as a function of capillary number.
}
\label{fig-sketch-coalescence}
\end{figure}

Figure~\ref{fig-sketch-coalescence} shows the computational configuration and dynamic wetting characterization adopted for the simulations. 
Two identical surface-attached bubbles, each with an initial diameter of 1.4 mm, are placed on a solid substrate and allowed to merge through capillary bridge growth.
The simulations are performed using the numerical framework described in Section~\ref{method}, with air--water properties under ambient conditions. 
The dynamic wetting parameters for the TMCS-treated glass substrate are summarized in Table~\ref{tab-coal-CA35}. 
Gas dissolution and interfacial mass transfer are neglected owing to the short timescale of the coalescence process.
A minimum mesh size of approximately $\Delta_{\min}\approx2.5\,\mathrm{\mu m}$ is employed near the interface and contact line region. This resolution has previously been validated for coalescence-induced bubble dynamics \cite{bashkatov2025electrolyte} and is sufficient to accurately capture bridge evolution, bubble recoil, and contact line motion.

    \begin{table}[]
    \centering
    \caption{\label{tab-coal-CA35}Dynamic wetting parameters of a  TMCS-treated glass surface with $\theta_s=35^\circ$.}
    \begin{tabular}{llllll}
    \hline
             $\theta_s (\mathrm{^\circ})$    &$\theta_a  (\mathrm{^\circ})$  &$\theta_r  (\mathrm{^\circ})$ & $\xi  (\mathrm{Pa \cdot s})$ & \quad $C$ & \quad$\epsilon$ \\\hline
            35 & 65 & 25 & \quad 0.02 & $2\times 10^4$ & $1.0\times 10^4$    \\\hline
    \end{tabular}
    \end{table}

\begin{figure}
\centering
\includegraphics[width=0.99\textwidth]{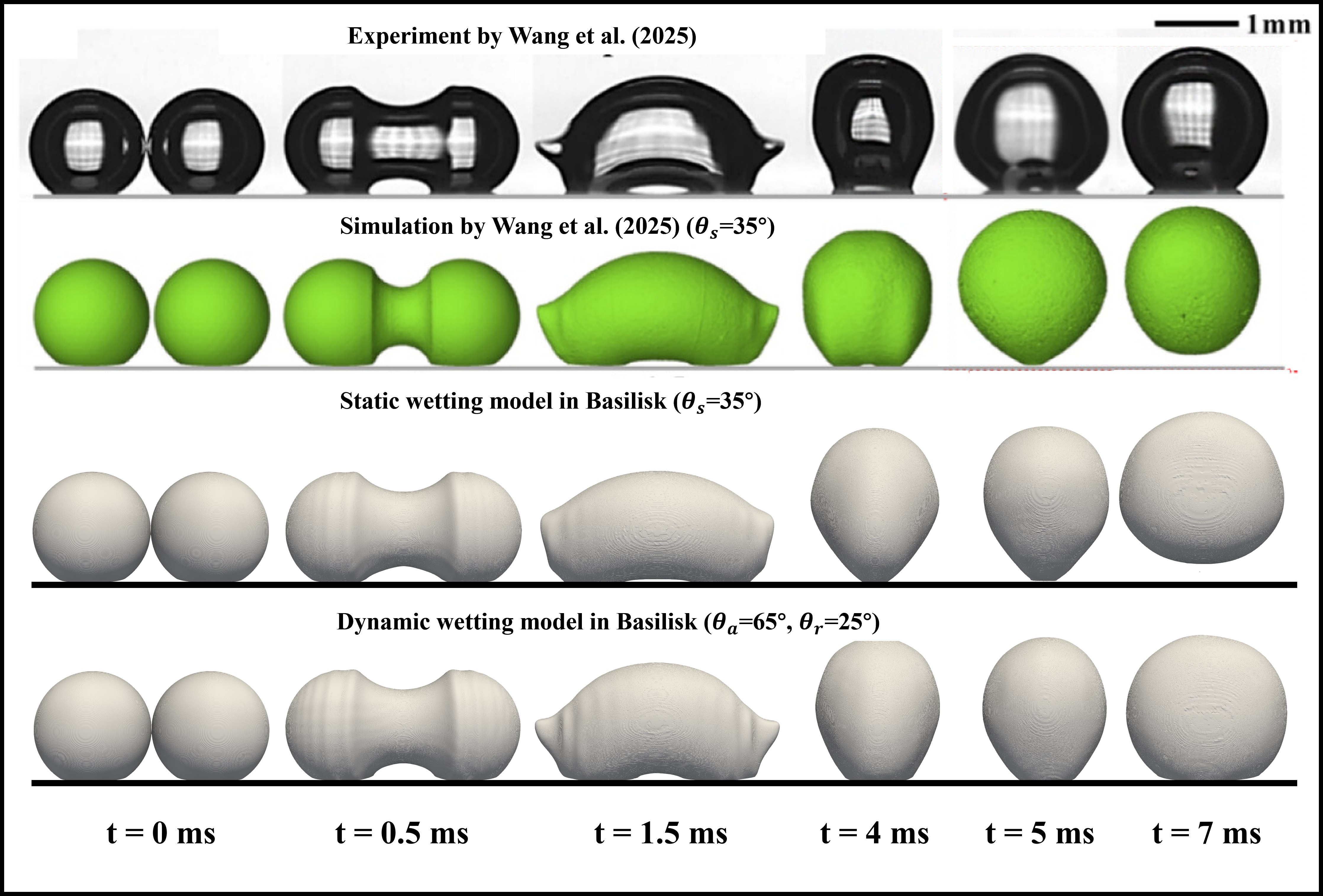}
\caption{\label{fig-valid-coalescence} Comparison of coalescence-induced detachment of two surface-attached bubbles obtained from experiments and numerical simulations.
The first row shows the experimental observations reported by Wang et al.~\cite{wang2025role}. 
The second row presents the numerical results of Wang et al.~\cite{wang2025role}. 
The third and fourth rows show the present simulations using static and dynamic wetting models, respectively. 
The dynamic wetting model incorporates contact-angle hysteresis ($\theta_a=65^\circ$, $\theta_r=25^\circ$, $\mathrm{CAH}=40^\circ$), whereas the static wetting model employs a constant equilibrium contact angle of $\theta_s=35^\circ$.}
\end{figure}

Figure~\ref{fig-valid-coalescence} compares the experimental observations reported by Wang et al.~\cite{wang2025role}, the numerical results presented in the same study, and the present simulations using both static and dynamic wetting models. 
Following contact, a liquid bridge rapidly develops between the two bubbles and expands due to capillary forces. 
The resulting capillary--inertial recoil generates significant interface deformation and drives the merged bubble upward. 
In the experiments, the bubble undergoes substantial shape oscillation but remains attached to the substrate throughout the observation period.
The static wetting models predict a markedly different evolution. 
Both the simulation of Wang et al.~\cite{wang2025role} and the present static-contact-angle model exhibit excessive contraction of the contact region during recoil, resulting in premature lift-off of the merged bubble. 
This behavior indicates that the mobility of the contact line is overestimated when contact-angle hysteresis is neglected.

In contrast, the present dynamic wetting model successfully reproduces the experimentally observed coalescence dynamics. The incorporation of contact-angle hysteresis introduces additional resistance to contact line motion, suppresses excessive contact line recession, and increases the threshold required for detachment. Consequently, the merged bubble remains attached to the wall after coalescence, in agreement with the experimental observations. Although some differences in the detailed bubble shape evolution remain, these discrepancies are expected because the advancing and receding contact angles, contact line friction coefficient, and hysteresis characteristics of the TMCS-treated surface were not reported in the original experiments and therefore cannot be prescribed exactly in the simulations. Nevertheless, the overall coalescence behavior and final attachment state are captured correctly. 
This finding is also consistent with the conclusions of Iwata et al.~\cite{iwata2022coalescing}, who showed that contact line dynamics and wall adhesion play a critical role in determining whether a coalescing bubble detaches from or remains attached to a solid surface.

\begin{figure}
\centering
\includegraphics[width=0.55\textwidth]{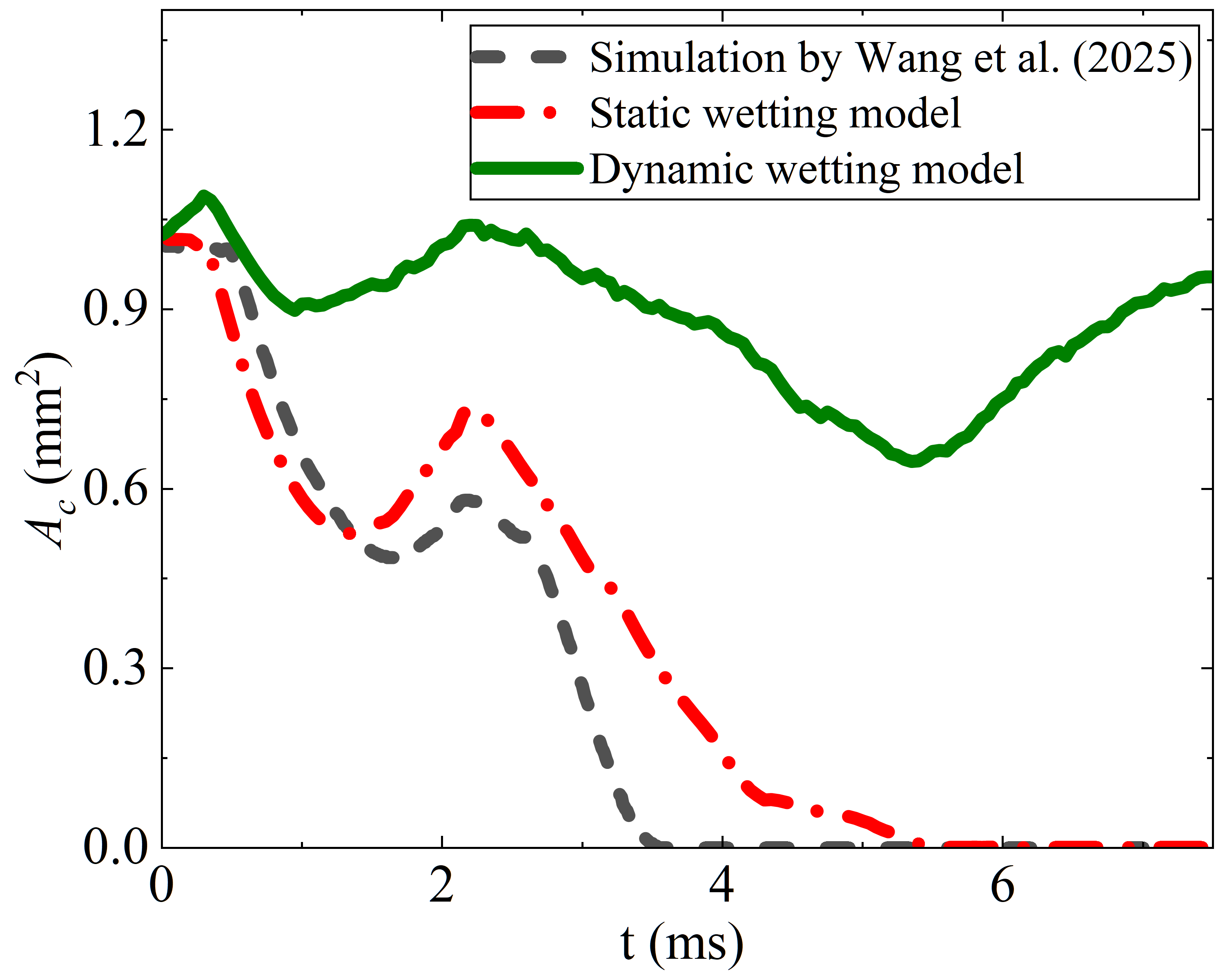}
\caption{\label{fig-area-coalescence} Time evolution of the gas--solid contact area $A_c$ during coalescence-induced detachment. Simulation by Wang et al.~\cite{wang2025role} (dashed line), the static wetting model (dash-dotted line), and the present dynamic wetting model (solid line).
}
\end{figure}

The physical mechanism underlying these differences is further quantified in Fig.~\ref{fig-area-coalescence}, which shows the temporal evolution of the gas--solid contact area $A_c$. 
For both the simulation of Wang et al.~\cite{wang2025role} and the static wetting model, the contact area decreases rapidly following bridge growth and capillary recoil. The contact area eventually vanishes within approximately $3$--$4\,\mathrm{ms}$, indicating complete loss of wall attachment and bubble detachment.
By contrast, the dynamic wetting model maintains a substantially larger contact area throughout the coalescence process. Although capillary recoil initially reduces the contact area, contact-angle hysteresis and contact line dissipation significantly slow the recession of the contact line and preserve a large fraction of the bubble footprint on the substrate. 
As a result, sufficient adhesion is maintained to prevent lift-off of the merged bubble. These results demonstrate that the evolution of the contact area is highly sensitive to dynamic wetting and directly reflects the competition between capillary recoil and contact line resistance.


\subsection{Bubble dynamics at surfaces under shear flow}
\label{results:shear}

Bubble transport under externally imposed shear flow represents another important class of wetting-controlled interfacial phenomena encountered in electrochemical systems, membrane reactors, and wall-bounded multiphase flows. In such systems, hydrodynamic shear induces asymmetric deformation of the bubble interface and contact region, while the resulting contact line motion strongly influences bubble sliding, migration, and eventual detachment from the surface~\cite{baczyzmalski2017growth,wang2022dynamic,uriarte2024phase}. This configuration therefore provides a useful model problem for examining the interplay between external forcing and dynamic wetting.

\begin{figure}
\centering
\begin{minipage}[b]{0.55\textwidth}
    \centering
    \includegraphics[width=\textwidth]{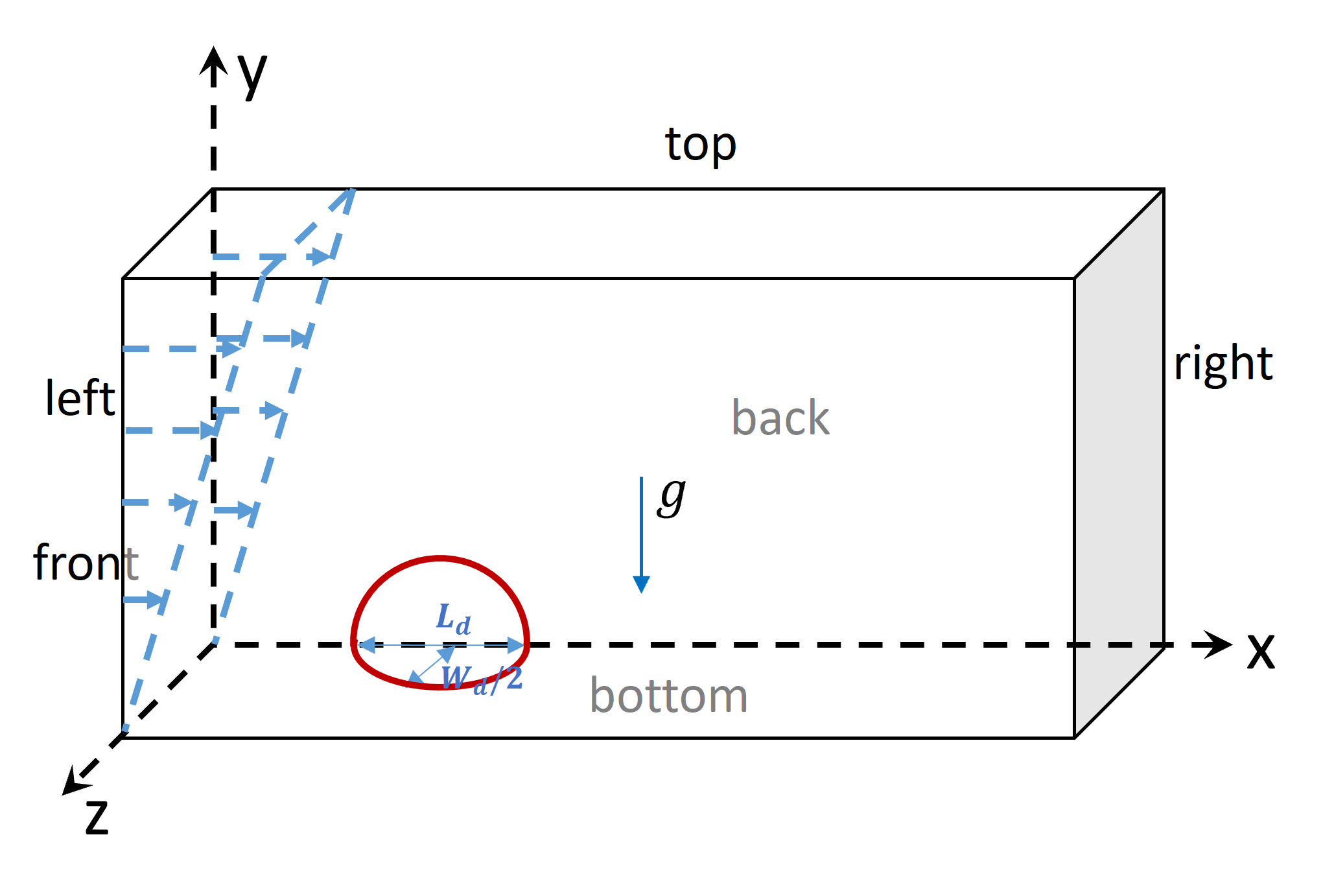}\\
    \textbf{(a)}
\end{minipage}
\hfill
\begin{minipage}[b]{0.43\textwidth}
    \centering
    \includegraphics[width=\textwidth]{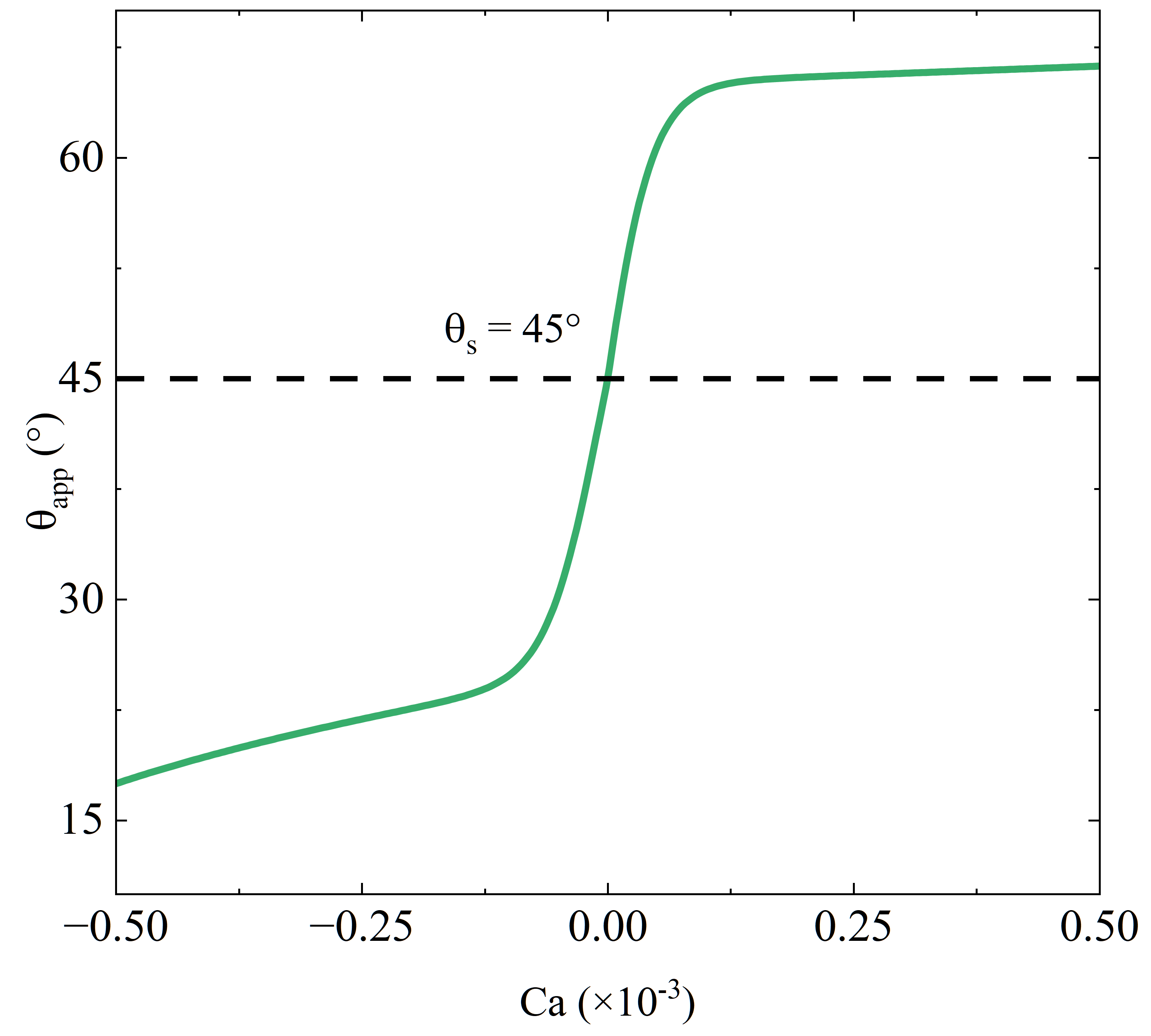}\\
    \textbf{(b)}
\end{minipage}
\caption{
 Numerical setup of a gas bubble in wall-bounded shear flow with dynamic wetting.
(a) Computational domain and initial configuration of two identical bubbles attached to a solid surface.
(b) Apparent contact angle as a function of capillary number.
}
\label{fig-sketch-bubble-shear}
\end{figure}

Figure~\ref{fig-sketch-bubble-shear} illustrates the computational configuration adopted to investigate bubble dynamics under shear flow. A surface-attached air bubble is placed on a hypothetical hydrophilic substrate and subjected to a linear shear flow. The fluid properties correspond to the air--water system at ambient conditions, while the dynamic wetting parameters are summarized in Table~\ref{tab-shear-dca}. The initial bubble is approximated as a spherical cap with radius $R_0=1,\mathrm{mm}$ and equilibrium contact angle $\theta_s=45^\circ$. A linear velocity profile $u_x=\dot{S}y$ is prescribed at the inlet to generate controlled shear forcing, and different shear rates $\dot{S}$ are considered to examine their influence on bubble deformation, sliding, and detachment. 
A minimum cell size of approximately $\Delta_{\min}\approx9.8\,\mu\mathrm{m}$ is employed throughout the simulations. This spatial resolution has been validated in our previous studies of dynamic wetting and bubble dynamics and is sufficient to resolve contact line motion, interfacial deformation, and the characteristic length scales associated with shear-induced bubble detachment.

    \begin{table}[]
    \centering
    \caption{\label{tab-shear-dca}Dynamic wetting parameters of a hypothetical hydrophilic surface.}
    \begin{tabular}{llllll}
    \hline
             $\theta_s (\mathrm{^\circ})$    &$\theta_a  (\mathrm{^\circ})$  &$\theta_r  (\mathrm{^\circ})$ & $\xi  (\mathrm{Pa \cdot s})$ & \quad $C$ & \quad$\epsilon$ \\\hline
            45 & 65 & 25 & \quad 0.02 & $2\times 10^4$ & $1.0\times 10^4$    \\\hline
    \end{tabular}
    \end{table}

\begin{figure}
\centering
\begin{minipage}[b]{0.435\textwidth}
    \centering
    \includegraphics[width=\textwidth]{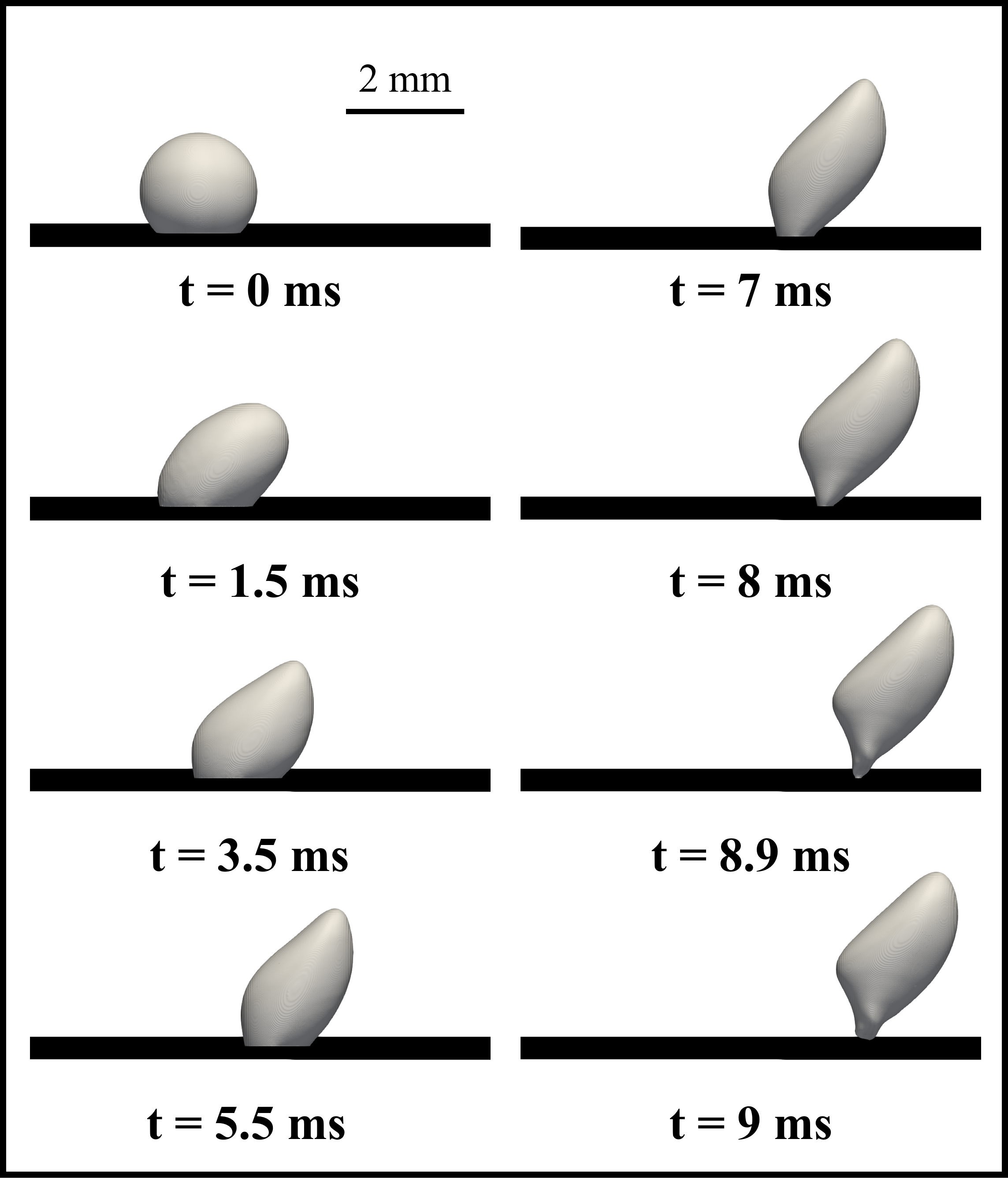}\\
    \textbf{(a)}
\end{minipage}
\hfill
\begin{minipage}[b]{0.49\textwidth}
    \centering
    \includegraphics[width=\textwidth]{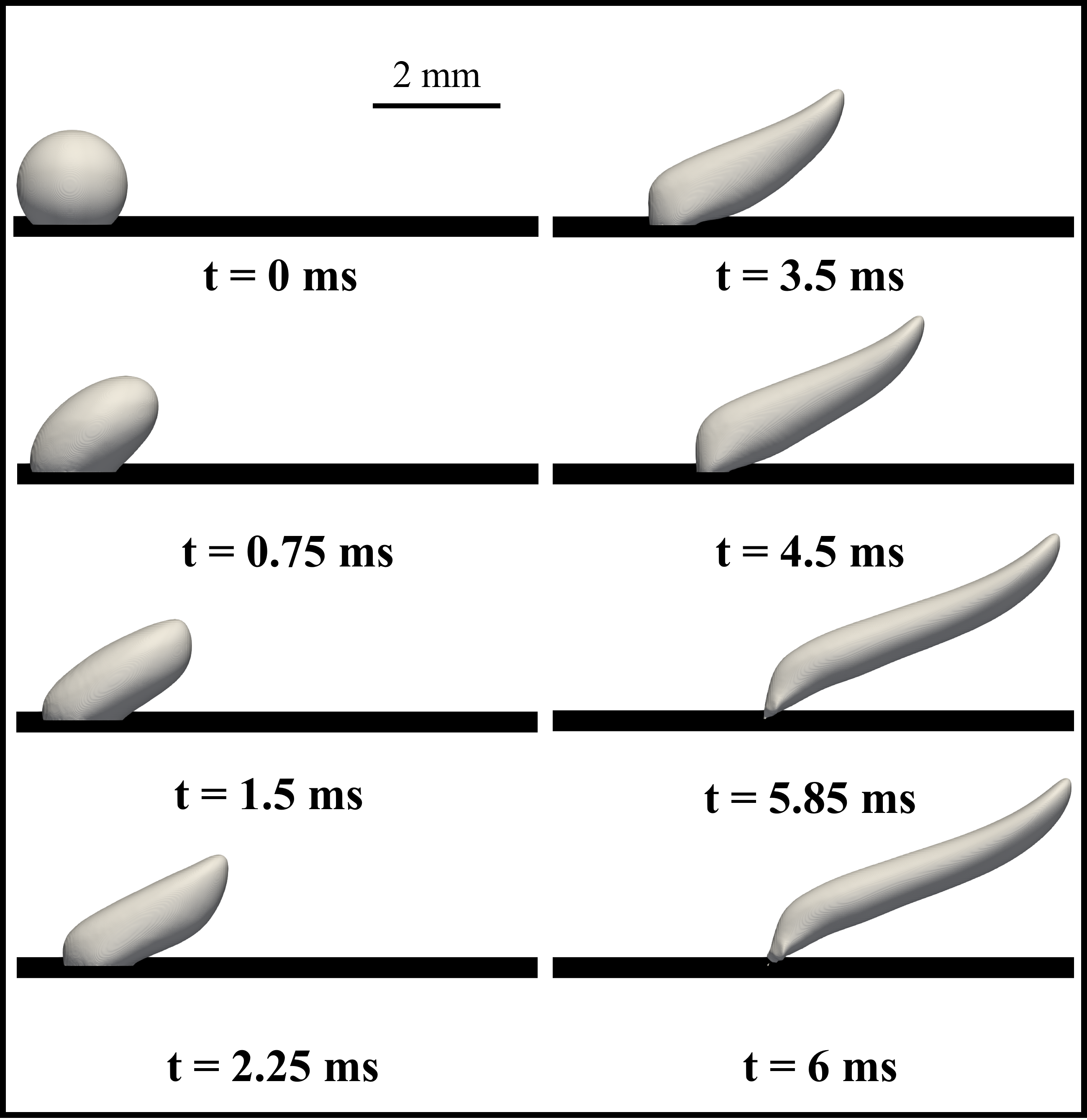}\\
    \textbf{(b)}
\end{minipage}
\caption{
 Time sequence of a wall-attached bubble subjected to shear flow at different shear rates.
 (a) $\dot{S}=400~\mathrm{s^{-1}}$.
 (b) $\dot{S}=800~\mathrm{s^{-1}}$.
 The snapshots (isosurfaces $f=0.5$ ) illustrate bubble deformation and sliding along the wall from $t=0$ to $t=9\,\mathrm{ms}$.
}
\label{fig-shear-flow}
\end{figure}

Figure~\ref{fig-shear-flow} illustrates the evolution of a wall-attached bubble subjected to shear flow. Two representative shear rates, $\dot{S}=400\,\mathrm{s^{-1}}$ and $\dot{S}=800\,\mathrm{s^{-1}}$, are considered to examine the influence of hydrodynamic forcing on bubble deformation, sliding, and detachment.
At $\dot{S}=400\,\mathrm{s^{-1}}$, the bubble initially deforms into an asymmetric shape and gradually elongates in the stream direction. The imposed shear generates a net hydrodynamic force that drives downstream migration of the contact region while maintaining attachment to the substrate. As shown in Fig.~\ref{fig-shear-flow}(a), the bubble undergoes progressive stretching and sliding during the first several milliseconds before finally detaching at approximately $t_d\approx9\,\mathrm{ms}$.
Increasing the shear rate to $\dot{S}=800\,\mathrm{s^{-1}}$ significantly accelerates the deformation process. The bubble develops a pronounced tail-like structure aligned with the flow direction and experiences stronger streamwise stretching. As shown in Fig.~\ref{fig-shear-flow}(b), the enhanced hydrodynamic loading promotes earlier destabilization of the attached state, resulting in detachment at approximately $t_d\approx6\,\mathrm{ms}$.

\begin{figure}
\centering
\includegraphics[width=0.8\textwidth]{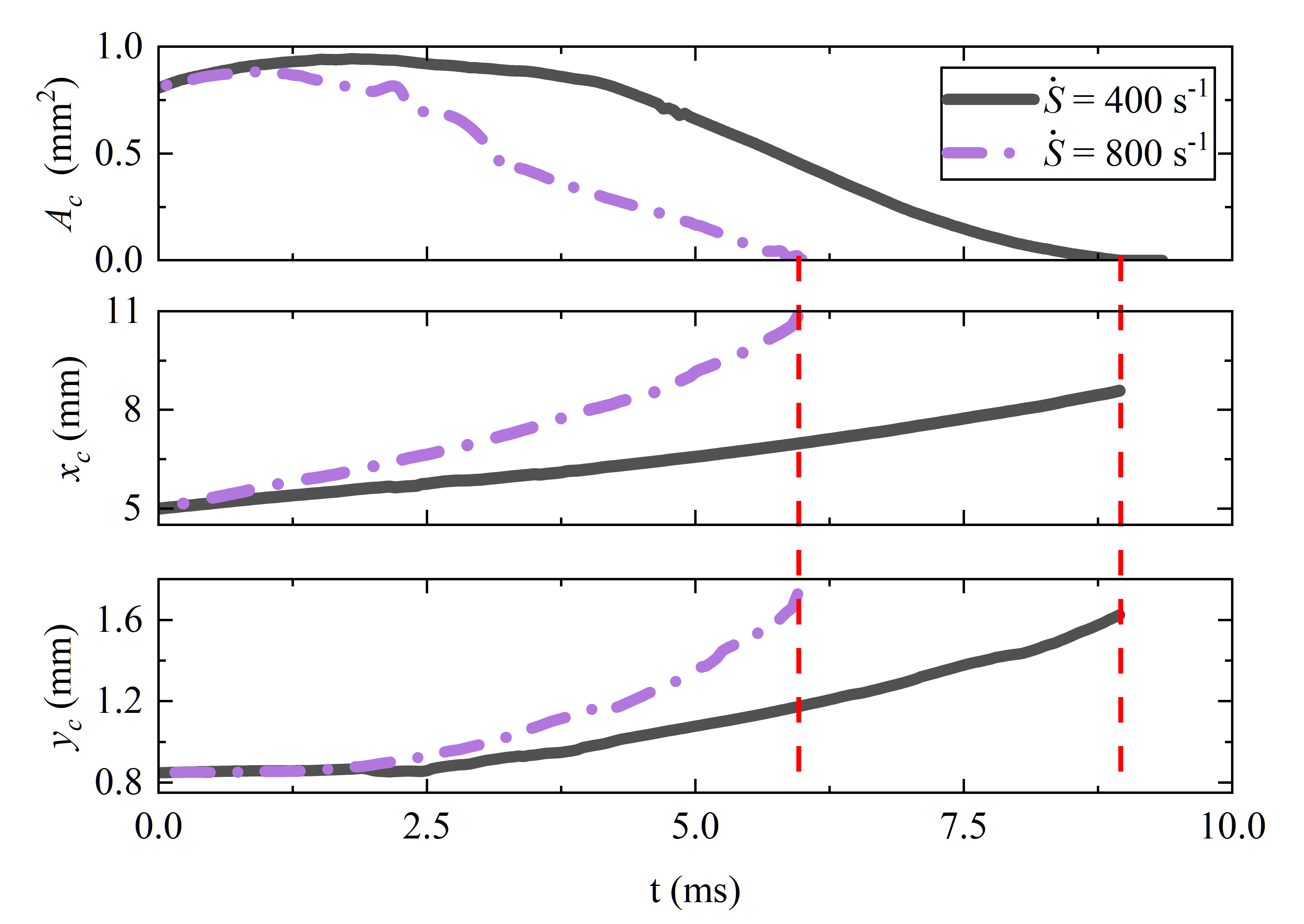}
\caption{\label{fig-area-shear} 
Temporal evolution of the contact area and centroid position of the bubble footprint under different shear rates.
(a) Contact area $A_c$.
(b) Streamwise position of the contact-area centroid $x_c$.
(c) Wall-normal position of the contact-area centroid $y_c$.
The dashed vertical lines indicate the bubble detachment times for the two shear-flow conditions.
}
\end{figure}

The influence of the shear rate is quantified in Fig.~\ref{fig-area-shear}, which shows the evolution of the contact area $A_c$ together with the centroid coordinates of the contact region, $x_c$ and $y_c$. For both cases, the contact area continuously decreases as the bubble deforms and the contact line recedes. However, the reduction is substantially faster at $\dot{S}=800\,\mathrm{s^{-1}}$, indicating an accelerated loss of wall adhesion under stronger shear forcing. Similarly, the centroid position in stream $x_c$ increases more rapidly, demonstrating enhanced bubble sliding along the substrate.
The normal position of the centroid on the wall $y_c$ also exhibits a pronounced increase prior to detachment. 
The faster growth of $y_c$ at $\dot{S}=800\,\mathrm{s^{-1}}$ reflects stronger lifting of the contact region and earlier onset of detachment. The vertical dashed lines in Fig.~\ref{fig-area-shear} indicate the detachment times for the two cases and clearly show that increasing shear rate substantially shortens the residence time of the bubble on the surface.
These results demonstrate that dynamic wetting governs the response of the contact region to external shear forcing. The coupled evolution of contact-area recession, streamwise migration, and wall-normal lifting determines the detachment process, while increasing shear rate accelerates all three mechanisms and promotes earlier bubble departure from the surface.

\section{Conclusion}
\label{conclu}

In this work, the influence of dynamic wetting on droplet and bubble dynamics at solid surfaces was investigated using a three-dimensional VOF framework coupled with a dynamic wetting model. 
The numerical method employs a geometrical reconstruction of the contact line velocity together with a dynamic contact angle formulation incorporating contact angle hysteresis, enabling robust simulation of moving contact lines in complex three-dimensional multiphase flows. Particular emphasis has been placed on wetting-controlled phenomena that involve rapid contact line motion, strong interface deformation, and topology changes, providing a stringent assessment of the dynamic wetting framework under highly transient conditions.

Three representative wetting-controlled interfacial flow problems were considered, namely droplet splashing, coalescence-induced bubble dynamics at solid surfaces, and bubble deformation and detachment under shear flow. 
For droplet splashing, the model successfully reproduced the experimentally observed spreading dynamics, rim instabilities, and crown-like splashing morphology, while providing improved predictions of the contact line evolution compared to existing dynamic contact-angle implementations. 
For bubble coalescence, the dynamic wetting model accurately captured the experimentally observed attachment behavior after coalescence and demonstrated that contact-angle hysteresis plays a key role in regulating the evolution of contact areas and suppressing premature detachment. 
Under externally imposed shear flow, the simulations showed that dynamic wetting strongly influences bubble deformation, sliding, and detachment through the coupled evolution of the contact region and moving contact line.

Despite the different physical configurations, all three cases reveal a common mechanism: dynamic wetting modifies the balance between capillary, viscous, inertial, and external forcing effects through its control of contact line motion. The resulting changes in contact line dissipation, hysteresis, and pinning/depinning behavior significantly affect interfacial deformation, topology evolution, and detachment dynamics.

The present study demonstrates that an accurate representation of dynamic wetting is essential for predictive simulations of wetting-controlled multiphase flows involving strong interface deformation and transient contact line motion. 
The proposed framework provides a versatile numerical tool for investigating a broad range of droplet and bubble transport processes relevant to electrochemical systems, boiling, and microfluidics.

\section{Acknowledgements} 
This work was supported by the Federal Ministry of Education and Research (BMBF) through the H2Giga-SINEWAVE project (Grant No.~03HY123E) and by the German Space Agency (DLR), funded by the Federal Ministry for Economic Affairs and Energy (BMWi), under Grant No.~DLR 50WM2058 (MADAGAS II).

\bibliographystyle{plain}
\bibliography{sn-bibliography}

\end{document}